\documentclass{pasj00}

\usepackage{threeparttable}

\newcommand{\fermi}{\textit{Fermi}}
\newcommand{\gr}{$\gamma$-ray}
\newcommand{\sax}{SAX J1808.4$-$3658}

\begin{document}
\SetRunningHead{Xing, Wang, Jithesh}{Possible \textit{Fermi} Detection of SAX J1808.4$-$3658}

\title{Possible \textit{Fermi} Detection of the Accreting Millisecond Pulsar
Binary SAX J1808.4$-$3658}

\author{Y. \textsc{Xing}, Z. Wang, and V. Jithesh}
\affil{Shanghai Astronomical Observatory, Chinese Academy of Sciences, 
80 Nandan Road, Shanghai 200030, China}
\email{wangzx@shao.ac.cn}


%


\maketitle

\begin{abstract}
We report the \textit{Fermi} Large Area Telescope (LAT) detection 
of a $\gamma$-ray
source at the position of SAX J1808.4$-$3658. This transient low-mass X-ray 
binary contains an accreting millisecond pular, which is only seen
during its month-long outbursts and 
likely switches to be rotation powered during its quiescent state. 
Emission from the $\gamma$-ray source can be described by a power law with
an exponential cutoff, the characteristic form for pulsar emission. 
Folding the source's 2.0--300\,GeV photons at the binary orbital period, 
a weak modulation is seen (with an H-test value of $\sim$17). 
In addition, three sets of archival \textit{XMM-Newton} data for
the source field are analyzed, and we
find only one X-ray source with 3--4$\sigma$ flux variations in the 2$\sigma$
error circle of the $\gamma$-ray source. However based on the X-ray properties, 
this X-ray source is not likely a background AGN, the major class 
of \fermi\ sources detected by LAT. These results support the possible 
association between the
$\gamma$-ray source and SAX J1808.4$-$3658 and thus the scenario that
the millisecond pulsar is rotation powered in the quiescent state. 
Considering a source distance of 3.5\,kpc for SAX J1808.4$-$3658,
the 0.1--300\,GeV luminosity is 5.7$\times 10^{33}$\,erg\,s$^{-1}$, implying
a \gr\ conversion efficiency of 63\% for the pulsar in this binary.
\end{abstract}

\section{INTRODUCTION}

Millisecond radio pulsars (MSPs) are formed from neutron star low-mass X-ray 
binaries (LMXBs; \cite{bv91}).
The primary neutron star in an LMXB can gain sufficient
angular momentum by accreting material from the companion through 
an accretion disk, and thus be `recycled' to a spin period of milliseconds.
The discovery of millisecond X-ray pulsations in the transient LMXB 
SAX J1808.4$-$3658 
has confirmed the formation scenario from the observational
side \citep{wv98,cm98}. Thus far over a dozen of so-called
accreting millisecond X-ray pulsars (AMXPs) have been found 
(\cite{pw12}). Nearly all of them are in transient systems, and
for these transients, X-ray pulsations are seen only during 
their X-ray outbursts.

Interestingly, it was suggested that several 
AMXPs actually switch to be rotation powered pulsars
during their quiescent states (\cite{bur+03}; \cite{wan+13} and 
references therein), although no direct evidence, such as pulsed radio
emission \citep{burgay+03}, was found over the time.
The recent observational identification of the MSP binary 
J1824$-$2452I in the globular cluster M28 has firmly confirmed 
the suggestion \citep{pap+13}. The binary was observed to have an X-ray
outburst, and during the outburst, the previously known radio MSP 
in the binary switched to appear like a typical AMXP. This confirmation 
has thus identified an interesting
feature for the evolution from LMXBs to MSP binaries, and we may suspect
that either these systems would probably be at the end of their LMXB phase or
it could be a common feature for the quiescent state of transient neutron 
star LMXBs (e.g., \cite{hei+15}).

The similar type of feature has also been seen in the recently identified
two transitional MSP binaries: J1023+0038 \citep{arc+09} and 
XSS~J12270$-$4859 (\cite{bas+14} and references therein). Extensive
observational studies have shown that they can switch between the states
of having an accretion disk and being disk free. One particular property
in them is that they have sufficiently bright $\gamma$-ray emission and
are detectable by the \textit{Fermi} Large Area Telescope (LAT; see
\cite{tam+10,sta+14,tak+14} for J1023+0038; and see
\cite{hil+11,mar+13,xw14} for XSS~J12270$-$4859).
The emission is variable, 
stronger in the active accretion state than that in the disk-free state 
\citep{sta+14,tak+14,xw14}. Given the property similarities between 
the two MSP binaries and AXMP binaries, it is thus highly possible that
AMXPs would also have significant $\gamma$-ray emission, as they stay
in their quiescent state and would be rotation-powered most time. 
This possibility has been explored by
\citet{xw13} by searching for $\gamma$-ray emission from four AMXP systems,
which include SAX~J1808.4$-$3658.
Nearly four year LAT data for the four AMXPs were analyzed, but 
no $\gamma$-ray emission was found. However, given the improved sensitivity
of \fermi\ over the last two years (see \S~\ref{subsec:si} for details), 
we re-analyzed the LAT data for them. We found significant $\gamma$-ray 
emission at the position consistent with that of SAX~J1808.4$-$3658.
Here in this paper, we report the results.

\section{Observation}   
\label{sec:obs}

LAT is a $\gamma$-ray imaging instrument onboard the \textit{Fermi} Gamma-ray
Space Telescope.  It makes all-sky survey 
in an energy range from 20 MeV to 300 GeV \citep{atw+09}. In our analysis,
we selected LAT events from the \textit{Fermi} Pass 7 Reprocessed (P7REP) 
database inside a $\mathrm{20^{o}\times20^{o}}$ region centered 
at the optical position of \sax, which is 
R.A.=18$^{\rm h}$08$^{\rm m}$27$\farcs$62, 
Decl.=$-$36$^{\circ}$58$'$43$\farcs$3 (equinox J2000.0) obtained 
in \citet{har+08}. 
We kept events during the time period from 2008-08-04 15:43:36 (UTC) to 
2014-11-10 21:04:57 (UTC) and in the energy range of 100\,MeV to 300\,GeV. 
In addition, only events with zenith angle less than 100\,deg  
and during good time intervals were selected. The former prevents 
the Earth's limb contamination, and for the latter, the quality of 
the data was not affected by the spacecraft events. 

\section{Data Analysis and Results} 
\label{sec:ana}

\subsection{Source Identification}
\label{subsec:si}

We included all sources within 20 deg in the \textit{Fermi} second 
source catalog \citep{nol+12} centered at the position of 
\sax\ \citep{har+08} to make the source model. The spectral function 
forms of the sources are provided in the catalog. 
The spectral parameters of the sources 
within 5 deg from \sax\ were set free, and 
all other parameters of the sources were fixed at their catalog values. 
A point source at the optical position of \sax\ was also included in 
the source model, with its emission modeled by a simple power law.
In addition, we used the spectrum model gll\_iem\_v05\_rev1.fits and 
the spectrum 
file iso\_source\_v05.txt for the Galactic and the extragalactic diffuse 
emission, respectively, in the source model. 
The normalizations of the diffuse components were set as free parameters.

Using the LAT science tools software package {\tt v9r33p0}, we performed
standard binned likelihood analysis to the LAT data.
Events below 200 MeV were rejected
because of the relative large uncertainties of the instrument response 
function of the LAT in the low energy range. 
Energy ranges of 0.2--300, 0.5--300, 1--300, and 2--300 GeV were tested
in the analysis.
A source at the optical position was detected with Test Statistic (TS) values 
of 32, 34, 26, and 31, respectively. 
The TS value at a specific position, calculated from 
TS$= -2\log(L_{0}/L_{1})$ (where $L_{0}$ 
and $L_{1}$ are the maximum likelihood values for a model without and with 
an additional source respectively), is a
measurement of the fit improvement for including the source, and
is approximately the square of the detection significance 
of the source \citep{abd+10}. 
Thus the source was best detected in 0.5--300\,GeV with a significance 
of $\simeq$5$\sigma$.
We extracted the 
TS maps of a $2\arcdeg\times 2\arcdeg$ region centered 
at the position of \sax\ in the four energy ranges, with all 
sources in the source model considered except the source we found.
No catalog sources are within the square region.
In Figure~\ref{fig:tsmap}, the 0.5--300\,GeV TS map is shown.

We ran \textit{gtfindsrc} in the LAT software package to determine
the position of the source using photons in 0.5--300 GeV, 
and obtained 
the best-fit position R.A. = 272\fdg15, Decl. = $-$37\fdg06 
(equinox J2000.0), with 1$\sigma$ nominal uncertainty of 0\fdg05. 
The 2$\sigma$ error circle is marked in Figure~\ref{fig:tsmap} as 
a dark dashed circle. 
The optical position of \sax\ (mark by a dark cross in Figure~\ref{fig:tsmap}) 
is 0\fdg08 from the best-fit position and within the 2$\sigma$ error circle, 
suggesting possible association of the \gr\ source with \sax. 
Below we considered the source as the candidate \gr\ counterpart to \sax.

In our TS maps, separate excess \gr\ emission at the northwest corner appears
(Figure~\ref{fig:tsmap}).
We investigated whether it possibly contaminated our detection
of the candidate counterpart. We found that it is
consistent with being a point source at a position of
R.A. = 271\fdg4, Decl. = $-$36\fdg4 (equinox J2000.0; 
1$\sigma$ nominal uncertainty is 0\fdg1).
Including this source in our source model,
it can be totally removed from the TS maps, and the results of the position 
and spectrum (see Section~\ref{subsec:sa}) of the counterpart source did 
not have significantly differences (consistent within uncertainties).

In addition, since the source position is toward
the Galactic center direction ($G_b\simeq -8\fdg1$),
we also checked if the uncertainty on the Galactic diffuse emission could
produce false detection of the \gr\ source.
We manually increased the normalization of the Galactic diffuse component 
to a value 5$\sigma$ above the best-fit value, 
the $>$0.5 GeV \gr\ emission at the position of \sax\ was still detected,
with TS$\simeq$32. 
We found only when we increased the Galactic diffuse emission by 10\%
(approximately 77$\sigma$ above the best-fit value), 
the $>$0.5 GeV emission was then detected with TS$\simeq$9 (i.e., 
$\sim$3$\sigma$ detection significance).

The \gr\ source was not detected in the previous search 
using nearly four-year LAT data in $>$0.2\,GeV energy range \citep{xw13}. 
We re-analyzed the LAT data in the same time interval from 
2008-08-04 15:43:36 (UTC) to 2012 July 8 18:59:57 (UTC), and
obtained a TS value of $\simeq$25 at the position of \sax,
which is much higher than the value of $\sim$3 previously reported. 
Therefore the detection is due to the improved sensitivity of 
the \fermi\ telescope. The database used in \citet{xw13} is \fermi\ Pass 7, 
comparing to Pass 7 Reprocessed in this work. The LAT science tools software 
package and the Instrument Response Functions (IRFs) have also been updated 
from {\tt v9r27p1} to the current {\tt v9r33p0}
and from P7SOURCE\_V6 to P7REP\_SOURCE\_V15, respectively.
In addition, the diffuse emission models have also been updated. 
All these changes have improved the point source detection sensitivity 
of the \fermi/LAT\footnote{http://www.slac.stanford.edu/exp/glast/groups/canda/lat\_Performance.htm}.

\subsection{Spectral Analysis}
\label{subsec:sa}

Including the \gr\ source in the source model, 
we performed standard binned likelihood analysis to the LAT data, 
with emission of this source modeled with an exponentially cutoff power law,
which is characteristic of pulsars \citep{abd+13}.
In addition a simple power law was also used.
We used data in $>$0.2 GeV energy range to obtain a overall description 
of the \gr\ spectrum of the source.
A photon index of $\Gamma$ = 2.2$\pm$0.1 with a TS$_{pl}$ value of $\sim$32 
was obtained for the power-law model, and a photon index of
$\Gamma$ = 1.6$\pm$0.4 and a cutoff energy of E$_{c}$ = 5.5$\pm$3.7 GeV 
with a TS$_{exp}$ value of $\sim$37 were obtained for the exponentially 
cutoff power-law model. The low energy cutoff was thus detected with 
$>$2$\sigma$ significance (estimated from $\sqrt{{\rm TS}_{cutoff}}$, where 
TS$_{cutoff}\simeq {\rm TS}_{exp}-{\rm TS}_{pl}\simeq 5$; \cite{abd+13}). 
While the significance is low, this result also favors the possible 
association of the $\gamma$-ray source with \sax. 
These spectral results are summarized in Table~\ref{tab:likelihood}.

\begin{figure}
\centering
\includegraphics[width=3.4in]{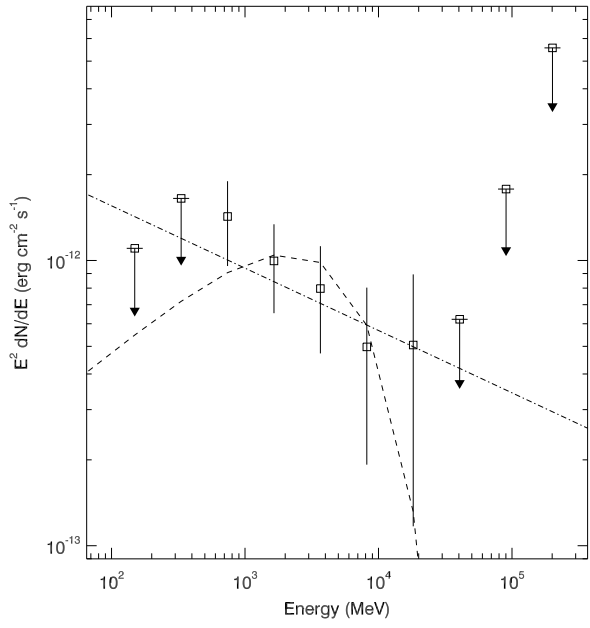}
\caption{\gr\ spectrum of \sax. The exponentially cutoff power-law 
and the power-law fits obtained from maximum likelihood analysis  
are shown as the dashed curve and dot-dashed line, respectively.}
\label{fig:spectrum}
\end{figure}

We then extracted the \gr\ spectrum for the source, by considering 
the emission as a point source with a power-law spectrum at 
the optical position of \sax\ and performing maximum likelihood analysis 
to the LAT data in 10 evenly divided energy bands in logarithm 
from 0.1--300 GeV. The photon index was fixed at 2.2. Only 
spectral points with TS$\geq$4 were kept, and the 95\% upper limits in 
other energy bins were derived. The spectrum extracted by this method is 
less model dependent and provides a more detailed description for 
the \gr\ emission of the source.
The obtained spectrum is shown in Figure~\ref{fig:spectrum}, 
and the energy flux values are given in Table~\ref{tab:spectrum-point}. 
It can be seen that the exponentially cutoff power law fits the data better,
particularly in low energy ranges where no \gr\ emission was significantly 
detected.

\subsection{Variability Analysis}
\label{subsec:ta}

We performed timing analysis to the LAT data of the \sax\ region to search 
for possible \gr\ pulsations. We folded the LAT data according to 
the X-ray ephemeris given in \citet{har+09}. The optical position of 
\sax\ was used for the barycentric corrections to photon arrival times, 
and photons within $R_{max}$ ($R_{max}$ ranges from 0\fdg1--1\fdg0 with 
a step of 0\fdg1) from the position were collected. 
Different energy ranges ($>$0.2, $>$0.5, $>$1, and $>$2 GeV)
were tested in folding. No pulsation signals were detected, and the
H-test values were smaller than 9 (corresponding to $<$3$\sigma$ detection 
significance; \cite{jrs89}). 

\begin{figure}
\centering
\includegraphics[width=3.4in]{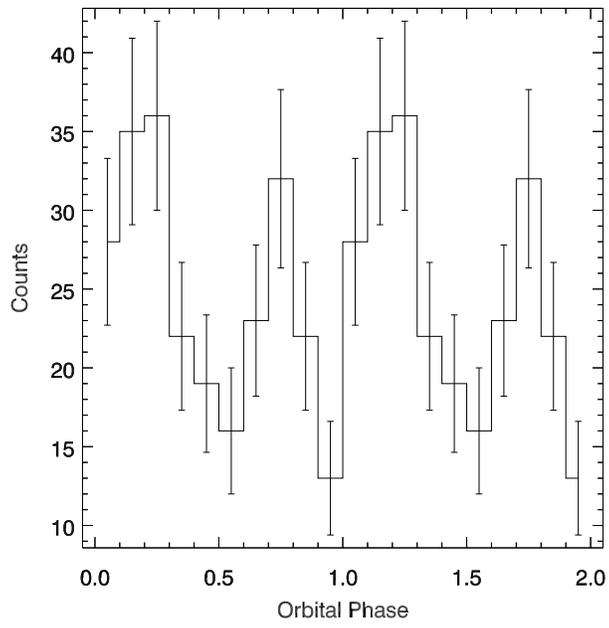}
\caption{2--300 GeV light curve folded with using the X-ray orbital 
parameters \citep{har+09}. The phase 
zero is at the ascending node of the MSP in \sax.}
\label{fig:timing}
\end{figure}

We folded the LAT data using the orbital parameters given 
in \citet{har+09}. We found that the highest orbital signal was revealed 
in the $>$2 GeV energy range using photons within 0\fdg6 from the optical 
position of \sax. The folded light curve, which has
an H-test value of $\sim$17 (corresponding to $>$3$\sigma$ detection 
significance, \cite{jrs89}), is shown in Figure~\ref{fig:timing}. The phase 
zero is set at the ascending node of the pulsar in \sax.

We also obtained the light curves for the \gr\ source, with different
time intervals (e.g., 30, 100, and 300 days) used. Due to the faintness 
of the source, no significant flux variations can be determined from
the light curves.

\subsection{\textit{XMM-Newton} data Analysis}

We searched in the SIMBAD Astronomical Database. Most sources
identified in the \gr\ source region are not high-energy but star-type 
objects. There are
two other X-ray sources, SAX J1808.5$-$3703 and SAX J1809.0$-$3659,
previously reported in the region \citep{wij+02,cam+02}.
While the two sources are 0\fdg02 and 0\fdg11 away from our best-fit 
position, respectively, there are also other X-ray sources detected in
the region but not studied in detail \citep{cam+02}.

We thus searched and found three archival {\it XMM-Newton} observations
of SAX J1808.4$-$3658 available. They were carried out on 2001 Mar. 24 
(ObsID : 0064940101; exposure 39.5 ks), 2006 Sept. 15 
(ObsID : 0400230401; exposure 55.1 ks), and 
2007 Mar. 10 (ObsID : 0400230501; exposure 57.8 ks). 
We analyzed the European Photon Imaging Camera (EPIC) pn and MOS data 
using the standard tools of the XMM-Newton Science Analysis Software 
(SAS, version 14.0). For the first observation, pn was used in the timing
mode and the data were not included in our analysis.
We excluded the high particle flaring background by creating the good time 
intervals (GTI) based on the count rate cut-off criteria. We extracted 
the full-field background light curve in the 10-12 keV band and selected 
the GTI with count rate $< 0.8$ and $< 0.4$\,cts\,s$^{-1}$ for pn and 
MOS data respectively. The data were then filtered to the good X-ray events 
(FLAG == 0) with PATTERN $\le$ 4 for pn and PATTERN $\le$ 12 for MOS in 
the 0.3--10 keV energy band. 

\begin{figure}
\centering
\includegraphics[width=3.4in]{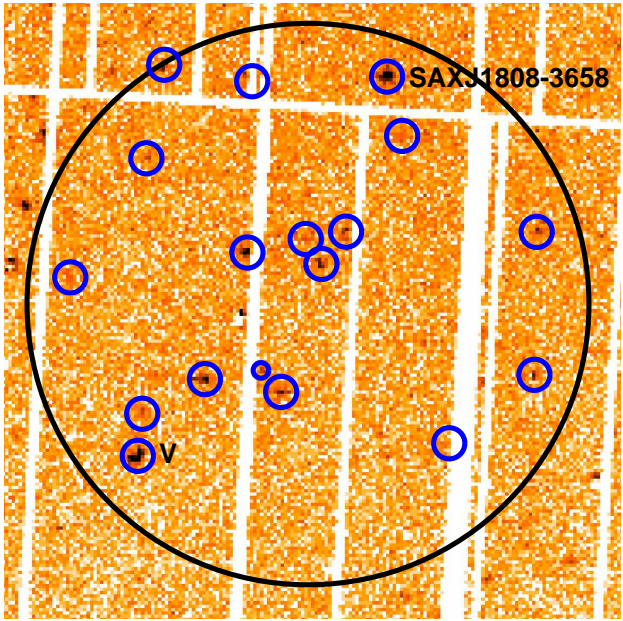}
\caption{\textit{XMM-Newton} pn image of the \sax\ field. The large circle
indicates the 2$\sigma$ error circle of the \fermi\ \gr\ source, in
which 17 X-ray sources were detected (marked with small blue circles).
The source found with 3--4$\sigma$ flux variations is marked with $V$.}
\label{fig:xray}
\end{figure}

We then performed the source detection routine ({\tt EDETECT\_CHAIN}) 
on 2007 EPIC-pn data and identified 17 field X-ray sources in the 2$\sigma$
\fermi\  error circle (radius of 0\fdg1) other than 
the AMXP, SAXJ 1808.4$-3$658. Their positions were astrometrically
calibrated by correlating the sources detected in the whole pn field with
the USNO B1.0 optical catalog (\cite{mon+03}; the SAS task {\tt EPOSCORR}
was used).  The X-ray field is shown in 
Figure~\ref{fig:xray}. We obtained the source counts of the 17 sources
in both pn and MOS data of the three observations. They were faint with
pn count rates of 0.5--7.9$\times 10^{-3}$ cts\,s$^{-1}$ and MOS count rates
of 0.1--2.6$\times 10^{-3}$ cts\,s$^{-1}$.
Comparing their count rates in the three observations, we investigated 
their variability. Out of 17 sources, 16 of them were 
non-variable ($< 3\sigma$).
Only one source, located at R.A.=18$^{\rm h}$08$^{\rm m}$54$\farcs$22, 
Decl.=$-$37$^{\circ}$06$'$50$\farcs$4 (equinox J2000.0; 1$\sigma$ positional
uncertainty is 0\farcs4), exhibited variability in the count rate at a 
significance level of $\sim 4\sigma$ and $\sim 3\sigma$ between 2006 and 2007 
pn data and 2001 and 2007 MOS data, respectively. However, this source did 
not show a significant variation ($\sim 2\sigma$) between the 2006 and 2007 
MOS data. 

We further studied this variable source by fitting its spectra with 
different models. The count rates were low in the 2001 and 2006 observations,
and probably due to this reason, an absorbed power law can provide a good
fit to the spectra. When the column density $N_H$ was set as a free parameter,
unphysically large power-law indices of 3--6 were favored. If we
fixed $N_H=1.3\times 10^{21}$\,cm$^{-2}$, the Galactic value toward the
source direction \citep{dl90}, the indices were lowered to 2.9--3.6, 
and the obtained MOS (absorbed) fluxes were in a range of
1.7--4.1$\times 10^{-14}$ erg\,cm$^{-2}$\,s$^{-1}$ and the 2006 pn flux 
was 1.4$^{+0.7}_{-0.5}\times 10^{-14}$.
However the 2007 pn spectrum, the most significantly detected (with
a count rate of 7.91$\pm 0.71\times 10^{-3}$\,cts\,s$^{-1}$) among the sources
and observations, can not be fit with a single model such as a power law 
(reduced $\chi^2>2$; 20 degrees of freedom). 
We searched the SIMBAD database
and USNO B1.0 catalog, no radio or optical counterparts (generally
down to 20 mag in $R$ band) were found.
Given the properties, we suggest that this source is either a low luminosity
X-ray binary or a background galaxy. The power law index values are too high
for an AGN (e.g., \cite{umu97}), the largest class among \fermi\ LAT sources
\citep{f3rd}.

\section{Discussion}
\label{sec:disc}

Carrying out maximum likelihood analysis of more than 6-year
\fermi\ $\gamma$-ray data of the source region of \sax, 
we have detected a \gr\ source with the best-fit position consistent with 
that of AMXP. The source's \gr\ spectrum can be
described by an exponentially cutoff power law.
The obtained parameters of $\Gamma$ = 1.6$\pm$0.4 and 
E$_{c}$ = 5.5$\pm$3.7 GeV are within the parameter ranges
for pulsars (0.4 $<\Gamma<$2, 0.4 GeV $< E_{c} <$ 5.9 GeV; see 
the \fermi\ second pulsar catalog, \cite{abd+13}), although 
the uncertainties are large due to low counts of the source. 
In addition, a possible orbital modulation has also been detected.
These results support the association of the \gr\ source with \sax. 

Observational studies of the transitional MSP binaries 
(i.e., J1023$+$0038 and XSS J12270$-$4859) have shown that 
\gr\ emission is brighter during their active state 
when an accretion disk appears \citep{sta+14,tak+14,xw14} than that
in the disk-free state,
while the radio pulsars are possibly still active but not observable. 
Very likely in the latter state, the \gr\ emission arises from
the magnetosphere of the MSPs (e.g., \cite{tak+14}).
In the former state, it has been suggested that
either the \gr\ emission is enhanced due to inverse Compton (IC) scattering 
of a cold pulsar wind off the optical/infrared photons from 
the accretion disk \citep{tak+14}, or alternatively self-synchrotron 
Compton processes at the magnetospheric 
region of a propellering neutron star is the possible working mechanism
for producing \gr\ emission \citep{ptl14}. 
Considering the similarities between the quiescent state of \sax\ and
the active state of the transitional MSP binaries, it is likely that
a same emission mechanism also works in the AMXP and thus the observed
\gr\ emission is expected.

The orbital modulation from this source was possibly detected. However 
because of the relative low significance and the unique modulation profile,
we can not draw a certain conclusion. The modulation has two brightness peaks 
around the inferior conjunction (phase 0.25, when the companion is in front 
of the neutron star) and the superior conjunction (phase 0.75, when 
the companion is behind the neutron star) respectively. Such modulation
has not been observed in other MSP binaries. Usually there is only one 
brightness peak, either around the inferior conjunction 
(see, e.g. PSR J1023$+$0038 in \cite{bog+11} and XSS J12270$-$4859 
in \cite{xw14}) and possibly due to the occultation of the photon emitting 
region by the companion, or around the superior conjunction (see, e.g., 
PSR B1957$+$20 in \cite{wu+12}, 2FGL J0523.3$-$2530 in \cite{xwn14}) and
possibly due to the viewing angle of the intrabinary interaction 
region \citep{wu+12,bed14}. Moreover, the orbital signals in these
MSP binaries are only seen when accretion disks are not present 
\citep{bog+11,bog+14,xw14}. Thus the possible orbital modulation
needs further confirmation from different studies.
Phase-resolved spectra may help identify the spectral differences and
the origin of the modulation. 
Unfortunately the photon counts from this \gr\ source were 
too low to allow such analysis.

Considering that the \gr\ emission is from \sax,  
the $>$0.1 GeV \gr\ luminosity of the source is 
$\sim 5.7d^{2}_{3.5} \times 10^{33}$\,erg\,s$^{-1}$ (for the exponentially 
cutoff power-law model) at the source distance of 3.5\,kpc \citep{gc06}. 
The spin-down luminosity $\dot{E}_{sd}$ of \sax\ is 
$\sim 9\times 10^{33}$\,erg\,s$^{-1}$ (\cite{har+09}), indicating a \gr\ 
conversion efficiency $\eta_{\gamma}$ of 63\%. 
The efficiency is above the `death line' defined in \citet{xw13} with 
the characteristic age of $\sim 12\times 10^{9}$\,yr 
for \sax\ (calculated from the pulsar parameters given in \cite{har+09}),
which supports the suggestion that older MSPs tend to have higher 
$\eta_{\gamma}$ values.

AGNs are the major class of \fermi\ LAT sources \citep{f3rd}, and they may
be identified from their strong variability (e.g., \cite{umu97,wil+14}).
We have analyzed three sets of \textit{XMM-Newton} X-ray imaging data of
the \sax\ field, and found 17 faint X-ray sources in the 2$\sigma$ error
circle of the \gr\ source. Among them, only one had 
3--4$\sigma$ low flux variations. However, this variable source did not 
have AGN-like emission, not
supporting that it could be a background AGN.
We caution that from the X-ray variability study,
none of the other 16 sources are likely an AGN, but we can not totally exclude 
the possibility. In order to identify their nature from spectral properties,
deep X-ray observations are needed.

\begin{figure}
\centering
\includegraphics[width=3.4in]{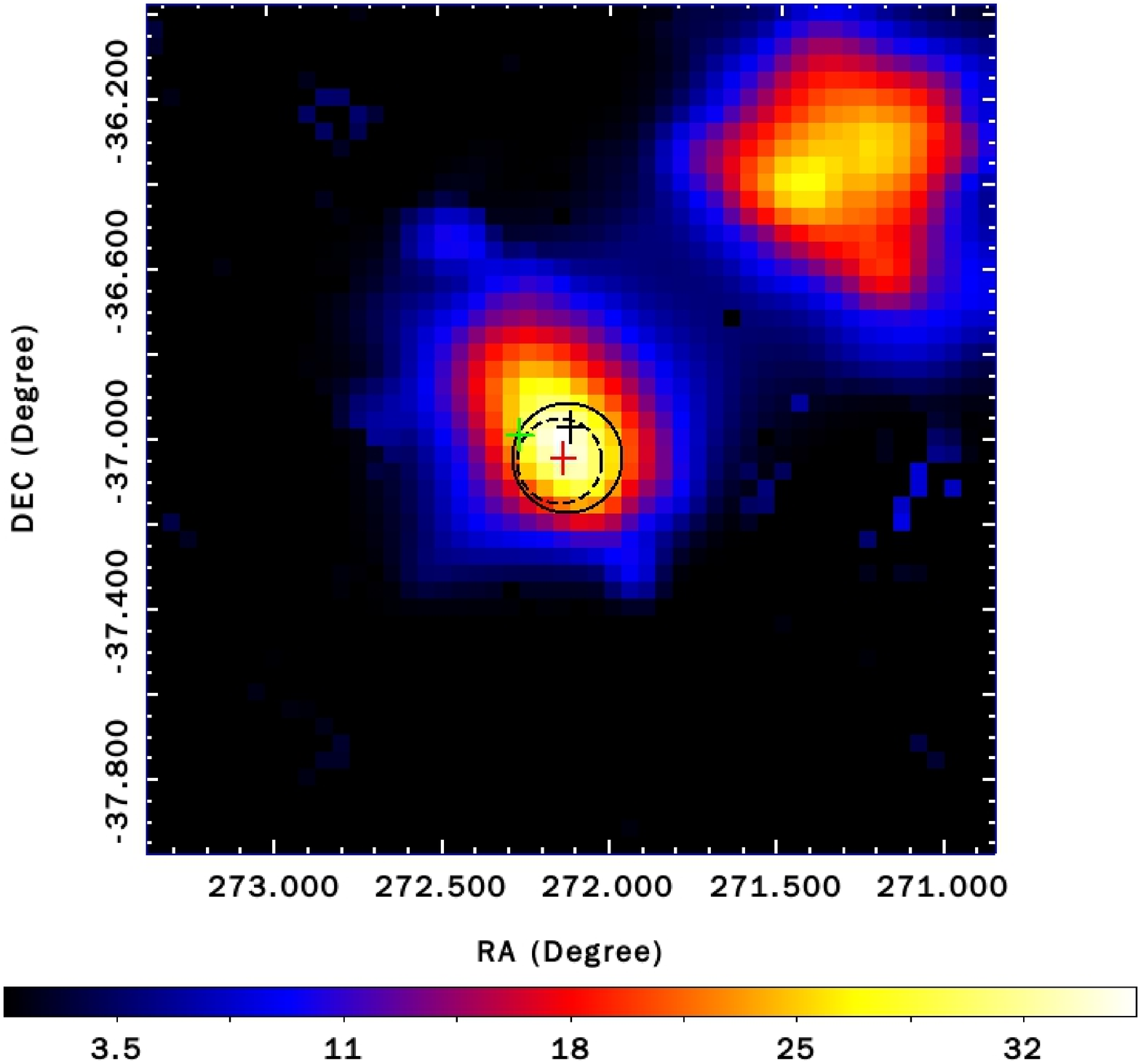}
\caption{TS map of a $\mathrm{2^{o}\times2^{o}}$ region,
with an image scale of 0.04\arcdeg\ pixel$^{-1}$,
centered at the position of \sax\ in $>$0.5 GeV energy range. 
The color bar indicates the TS values. 
The dark cross marks the optical position of \sax. The dark dashed and solid 
circles are the 2$\sigma$ error circles of 
the best-fit position obtained by us and given in the \textit{Fermi} 
third source catalog for 3FGL J1808.4$-$3703, respectively.}
\label{fig:tsmap}
\end{figure}

As we write the paper, the \textit{Fermi} third source 
catalog is released \citep{f3rd}, 
and we note that the source 
3FGL J1808.4$-$3703 is reported to be detected at the region of \sax. 
The catalog position of 3FGL J1808.4$-$3703 is R.A. = 272\fdg12, 
Decl. = $-$37\fdg05 (equinox J2000.0), consistent with the best-fit position 
we obtained within uncertainties (see Figure~\ref{fig:tsmap}).  
Thus our data analysis is confirmed by the catalog results.
The catalog source is also identified not to be unassociated with any known 
type of objects in the recently available catalogs. 

\bigskip
We thank Y. Tanaka and M. Gu for helpful discussion about AGN variability
and multiple energy properties.
This research made use of the High Performance Computing Resource in the Core
Facility for Advanced Research Computing at Shanghai Astronomical Observatory.
This research was supported by the Shanghai Natural Science 
Foundation for Youth (13ZR1464400), the National Natural Science Foundation
of China for Youth (11403075), the National Natural Science Foundation
of China (11373055), and the Strategic Priority Research Program
``The Emergence of Cosmological Structures" of the Chinese Academy
of Sciences (Grant No. XDB09000000). Z.W. is a Research Fellow of the
One-Hundred-Talents project of Chinese Academy of Sciences. J. V. acknowledges
the support by Chinese Academy of Sciences President's international
fellowship initiative (Grant No. 2015PM059).

\bibliographystyle{apj}

\clearpage
\begin{table*}
\begin{center}
\caption{Binned likelihood analysis results for \sax.}
\begin{threeparttable}
\begin{tabular}{lccccc}
\hline
\hline
Spectral model & $>$0.2 GeV Flux & $\Gamma$ & E$_{c}$ & TS \\
& (10$^{-9}$ photon cm$^{-2}$ s$^{-1}$) &  & (GeV) &   \\
\hline
Power law & 3.2 $\pm$ 0.9 & 2.2 $\pm$ 0.1 & ... & 32 \\
Power law with cutoff & 2.4 $\pm$ 0.9 & 1.6 $\pm$ 0.4 & 5.5 $\pm$ 3.7 & 37 \\
\hline
\end{tabular}
\end{threeparttable}
\label{tab:likelihood}
\end{center}
\end{table*}

\clearpage
\begin{table}
\begin{center}
\caption{Flux measurements of \sax.}
\begin{threeparttable}
\begin{tabular}{lcccc}
\hline
\hline
E & F$\mathrm{_{low}}$/10$^{-12}$ & TS \\
(GeV) & (erg cm$^{-2}$ s$^{-1}$) & \\
\hline
0.15 & 1.1 & 0 \\
0.33 & 1.7 & 0  \\
0.74 & 1.4 $\pm$ 0.5 & 10  \\
1.65 & 1.0 $\pm$ 0.3 & 10 \\
3.67 & 0.8 $\pm$ 0.3 & 9 \\
8.17 & 0.5 $\pm$ 0.3 & 5 \\
18.20 & 0.5 $\pm$ 0.4 & 4 \\
40.54 & 0.6 & 0 \\
90.27 & 1.8 & 0 \\
20.10 & 5.6 & 0 \\
\hline
\end{tabular}
\begin{tablenotes}
\footnotesize
\item[*] Columns 2 and 3 list the energy flux (E$^{2}$ $\times$ dN/dE) 
and the TS value in each energy bin, respectively. The fluxes without 
uncertainties are upper limits.
\label{tab:spectrum-point}
\end{tablenotes}
\end{threeparttable}
\end{center}
\end{table}

\clearpage

\clearpage

\clearpage

\clearpage

\end{document}